\documentclass[12pt,preprint]{aastex}

\usepackage{lineno}
\linenumbers

\usepackage{subfigure}

\usepackage[OT2,T1]{fontenc} 
\newlanguage\fakelanguage 
\newcommand\cyr{\fontencoding{OT2}\fontfamily{wncyr}\selectfont 
\language\fakelanguage} 
\DeclareTextFontCommand{\textcyr}{\cyr} 

\shorttitle{MW Campaign of PMN~J0948+0022}
\shortauthors{The \emph{Fermi}/LAT Collaboration}

\begin{document}

 
\title{Multiwavelength monitoring of the enigmatic\\Narrow-Line Seyfert 1 PMN J0948+0022 in March-July 2009}



\author{
A.~A.~Abdo\altaffilmark{2,3}, 
M.~Ackermann\altaffilmark{4}, 
M.~Ajello\altaffilmark{4}, 
M.~Axelsson\altaffilmark{5,6}, 
L.~Baldini\altaffilmark{7}, 
J.~Ballet\altaffilmark{8}, 
G.~Barbiellini\altaffilmark{9,10}, 
D.~Bastieri\altaffilmark{11,12}, 
B.~M.~Baughman\altaffilmark{13}, 
K.~Bechtol\altaffilmark{4}
% ...
}
% ...
\altaffiltext{2}{Space Science Division, Naval Research Laboratory, Washington, DC 20375, USA}
\altaffiltext{3}{National Research Council Research Associate, National Academy of Sciences, Washington, DC 20001, USA}
\altaffiltext{4}{W. W. Hansen Experimental Physics Laboratory, Kavli Institute for Particle Astrophysics and Cosmology, Department of Physics and SLAC National Accelerator Laboratory, Stanford University, Stanford, CA 94305, USA}
\altaffiltext{5}{Department of Astronomy, Stockholm University, SE-106 91 Stockholm, Sweden}
\altaffiltext{6}{The Oskar Klein Centre for Cosmoparticle Physics, AlbaNova, SE-106 91 Stockholm, Sweden}
\altaffiltext{7}{Istituto Nazionale di Fisica Nucleare, Sezione di Pisa, I-56127 Pisa, Italy}
\altaffiltext{8}{Laboratoire AIM, CEA-IRFU/CNRS/Universit\'e Paris Diderot, Service d'Astrophysique, CEA Saclay, 91191 Gif sur Yvette, France}
\altaffiltext{9}{Istituto Nazionale di Fisica Nucleare, Sezione di Trieste, I-34127 Trieste, Italy}
\altaffiltext{10}{Dipartimento di Fisica, Universit\`a di Trieste, I-34127 Trieste, Italy}
\altaffiltext{11}{Istituto Nazionale di Fisica Nucleare, Sezione di Padova, I-35131 Padova, Italy}
\altaffiltext{12}{Dipartimento di Fisica ``G. Galilei", Universit\`a di Padova, I-35131 Padova, Italy}
\altaffiltext{13}{Department of Physics, Center for Cosmology and Astro-Particle Physics, The Ohio State University, Columbus, OH 43210, USA}
% ...

\begin{abstract}
Following the recent discovery of $\gamma$ rays from the radio-loud narrow-line Seyfert 1 galaxy PMN J0948+0022 ($z=0.5846$), we started a multiwavelength campaign from radio to $\gamma$ rays, which was carried out between the end of March and the beginning of July 2009. The source displayed activity at all the observed wavelengths: a general decreasing trend from optical to $\gamma-$ray frequencies was followed by an increase of radio emission after less than two months from the peak of the $\gamma-$ray emission. The largest flux change, about a factor of about 4, occurred in the X-ray band. The smallest was at ultraviolet and near-infrared frequencies, where the rate of the detected photons dropped by a factor $1.6-1.9$. At optical wavelengths, where the sampling rate was the highest, it was possible to observe day-scale variability, with flux variations up to a factor of about 3. The behavior of PMN J0948+0022 observed in this campaign and the calculated power carried out by its jet in the form of protons, electrons, radiation and magnetic field are quite similar to that of blazars, specifically of flat-spectrum radio quasars. These results confirm the idea that radio-loud narrow-line Seyfert 1 galaxies host relativistic jets with power similar to that of average blazars. 
\end{abstract}

\keywords{quasars: individual (PMN J0948+0022) -- galaxies: active -- gamma rays: observations -- X-rays: galaxies -- ultraviolet: galaxies -- infrared: galaxies -- radio continuum: galaxies}

\section{Introduction}
The recent detection by \emph{Fermi Gamma-ray Space Telescope} of $\gamma$ rays from the radio-loud narrow-line Seyfert 1 galaxy (RL-NLS1) PMN J0948+0022\footnote{We note that the absolute magnitude of this source is $M_B=-23.6$, so formally matches also the definition of quasars.} ($z=0.5846$) opened new and interesting questions on the unified model of active galactic nuclei (AGN), the development of relativistic jets and the evolution of radio-loud AGN (Abdo et al. 2009a, Foschini et al. 2009a). Indeed, before \emph{Fermi}/LAT (Large Area Telescope) it was known that $\gamma$ rays from AGN are produced in blazars and radio galaxies, but we have to add also RL-NLS1s. 

NLS1s are active nuclei similar to Seyferts, where the optical permitted lines emitted from the broad-line region (BLR) are narrower than usual, with FWHM(H$\beta$)$<2000$~km~s$^{-1}$ (see Pogge 2000, for a review). Other characteristics are [OIII]/H$\beta<3$ and a bump of FeII, making them a peculiar class of AGN. NLS1s are different from Seyfert 2s, whose optical spectra typically display FWHM(H$\beta$)$<1000$~km~s$^{-1}$, [OIII]/H$\beta>3$ and no bump of FeII. NLS1s are also different from the naked AGN discovered by Hawkins (2004), a peculiar class of Seyferts without the BLR, which have [OIII]/H$\beta>>3$. Indeed, NLS1s do have both BLR and the narrow-line region (NLR), but the BLR emits only permitted lines narrower than in Seyfert 1s (Rodr\'iguez-Ardila et al. 2000).

NLS1s are generally radio-quiet, but a small fraction of them ($<7$\%, according to Komossa et al. 2006), are radio-loud. It is not clear how these sources fit into the framework of radio-loud AGN. Some studies of the average non-simultaneous multiwavelength properties (from radio to X-rays) of RL-NLS1s suggested some possibilities. Komossa et al. (2006) argued that RL-NLS1s could be some young stage of quasars, while Yuan et al. (2008) found some similarities to TeV BL Lacs, but having strong emission lines they would represent the so-called ``high-frequency peaked flat-spectrum radio quasars'' conjectured by Padovani (2007). Foschini et al. (2009b) found instead that there is no one-to-one correlation of RL-NLS1s properties with any specific type of blazar or radio galaxy. In some cases, there are similarities with flat-spectrum radio quasars, while others are like BL Lacs. 

Now, the first detection by \emph{Fermi}/LAT of $\gamma$ rays from one RL-NLS1 - namely PMN J0948+0022 - sets the definitive confirmation of the presence of a relativistic jet in these sources. The discovery of $\gamma-$ray emission from other sources of this type (Abdo et al., in preparation) raise RL-NLS1s to the rank of $\gamma-$ray emitting AGN. However, any average spectral energy distribution (SED) of a $\gamma-$ray loud AGN leaves open several important questions on the mechanisms of radiation emission, such as whether the synchrotron self-Compton (SSC) or the external Compton (EC) production mechanism is dominant at high-energies and where the zone is where most of the dissipation occurs. Because PMN J0948+0022 is the first object of this new class of $\gamma-$ray AGN, it is important to observe it for a long time, in order to understand if there is something unexpected and if its behavior is very different from blazars and radio galaxies or not.

With these aims in mind, we decided to set up a multiwavelength campaign on this source. The campaign involved several space and ground-based facilities across the whole electromagnetic spectrum, from radio to $\gamma$ rays (in alphabetical order): ATOM (Landessternwarte), F-GAMMA (Effelsberg), e-VLBI (EVN, LBA), \emph{Fermi}, G. Haro Telescope (INAOE), Mets\"ahovi, OVRO, RATAN-600, \emph{Swift}, SMARTS, MOJAVE (VLBA), WIRO. The period covered was between 2009 March 24 and July 5. We measured variability at multiple wavebands, modelled the resulting SEDs, and compared the results to those for more typical $\gamma-$ray blazars in the FSRQ and BL Lac classes. 

Throughout this work, we adopted a $\Lambda$CDM cosmology from the most recent \emph{WMAP} results, which give the following values for the cosmological parameters: $h = 0.71$, $\Omega_m = 0.27$, $\Omega_\Lambda = 0.73$ and with the Hubble-Lema\^{i}tre constant $H_0=100h$ km s$^{-1}$ Mpc$^{-1}$ (Komatsu et al. 2009). 

\section{Data Analysis}
\subsection{Fermi/LAT}
The data from the Large Area Telescope (LAT, Atwood et al. 2009) were analyzed using the same procedures outlined in Abdo et al. (2009a), but with a more recent version of the software ({\tt Science Tools v 9.15.2}), Instrument Response Function (\texttt{IRF P6\_V3\_DIFFUSE}, Rando et al. 2009) and background\footnote{Everything now publicly available at: \texttt{http://fermi.gsfc.nasa.gov/ssc/data/}}. Photons with energy above 100~MeV and between MJD 54910 (2009 March 20) and 55017 (2009 July 5) were selected. The quoted $1\sigma$ errors of the analyses are statistical only and systematic errors should be added. The most recent estimates set these values as 10\% at 100 MeV, 5\% at 500 MeV and 20\% at 10 GeV (Rando et al. 2009). 

The result of the fit with a power-law model in the form $F(E)\propto E^{-\Gamma}$ to the data integrated over the whole campaign gives an average flux ($E>100$~MeV) equal to $(1.5\pm 0.1)\times 10^{-7}$~ph~cm$^{-2}$~s$^{-1}$, a photon index $\Gamma = 2.48\pm 0.09$ with Test Statistic $TS=337$ (which is roughly equivalent to $18\sigma$, since $\sigma \sim \sqrt{TS}$; see Mattox et al. 1996 for the definition of TS). Comparison with the values obtained from the fit of the first 5 months of data, reported in Abdo et al. (2009a) and recalculated here as $F_{\rm E>100 MeV}=(1.6\pm 0.1)\times 10^{-7}$~ph~cm$^{-2}$~s$^{-1}$ with $\Gamma=2.7\pm 0.1$ ($TS = 386$), shows no changes in the average flux, but a slight spectral hardening during the present multiwavelength campaign. We observed no emission for energies above $\sim 2$~GeV.

\begin{figure}[!ht]
\centering
\includegraphics[angle=270,scale=0.7]{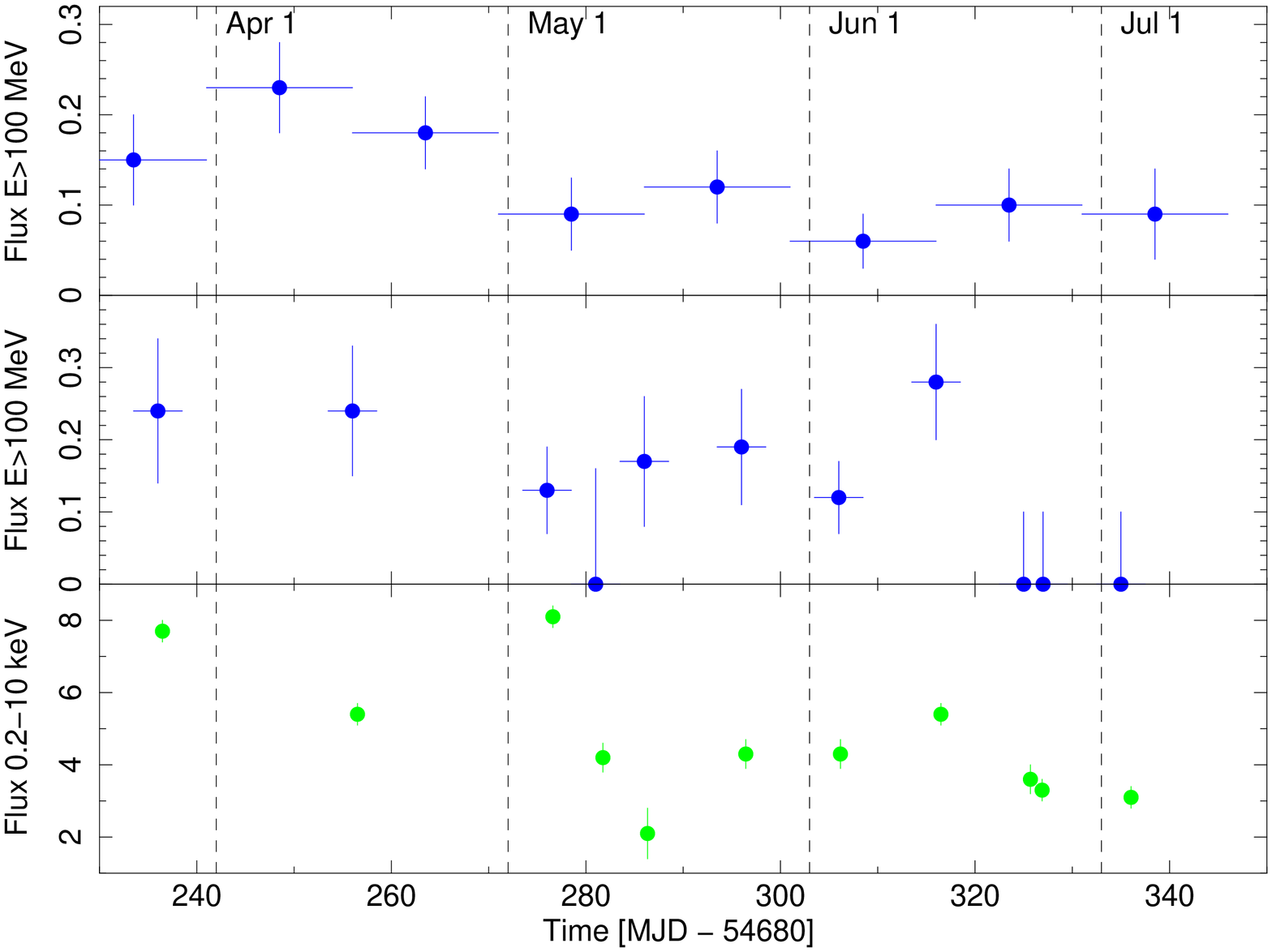}
\caption{(\emph{top panel}) $\gamma-$ray ($E>100$~MeV) light curve from \emph{Fermi}/LAT [$10^{-6}$~ph~cm$^{-2}$~s$^{-1}$], covering the whole period of the campaign. The bin time is $15$ days. (\emph{center panel}) $\gamma-$ray ($E>100$~MeV) light curve from \emph{Fermi}/LAT [$10^{-6}$~ph~cm$^{-2}$~s$^{-1}$], with $5$ days bin time centered on \emph{Swift} epochs.
(\emph{bottom panel}) X-ray ($0.2-10$~keV) light curves from \emph{Swift}/XRT [$10^{-12}$~erg~cm$^{-2}$~s$^{-1}$].}
\label{fig:latxrt}
\end{figure}

PMN J0948+0022 shows some variability on shorter timescales (Fig.~\ref{fig:latxrt}), but the weakness of the $\gamma-$ray flux hampers the study of the changes of its properties. Therefore, we decided to divide the campaign into three larger bins, by integrating and analyzing data month-by-month. The results are summarized in Table~\ref{tab:latsummary}. The better statistics allow us to measure a clear drop in flux of a factor $\sim 2$ from April to May and June, with a corresponding hardening of the spectral slope. We also note that the 2009 April flux was higher than the average flux in 2008 August-December. 

\subsection{Swift (BAT, XRT, UVOT)}
The \emph{Swift} satellite (Gehrels et al. 2004) observed PMN~J0948+0022 11 times, starting on MJD 54916.26 (2009 March 26 06:21 UTC) and ending on MJD 55015.53 (2009 July 3 12:41 UTC), with average exposures of $\approx 5$~ks for each observation. Data of BAT (Barthelmy et al. 2005), XRT (Burrows et al. 2005) and UVOT (Roming et al. 2005) have been analyzed by means of the \texttt{HEASoft~v.~6.6.3} software package, with default parameters (except as specified below) and the calibration database updated on 2009 June 5.

No detection was found with BAT in the hard X-ray energy band, after having integrated all the available data obtained in this campaign (total exposure $55.6$~ks, including the observation performed on 2008 December 5, see Abdo et al. 2009a), with an upper limit ($3\sigma$) of $3.2\times 10^{-10}$~erg~cm$^{-2}$~s$^{-1}$ in the $20-100$~keV energy band.

XRT was set to work in photon counting mode. Photons in the $0.2-10$~keV energy band and with grades 0-12 (single to quadruple pixels events) were selected. A check for pile-up gave negative results. The extracted spectrum was rebinned to have a minimum of $30$ counts per bin, in order to apply the $\chi^2$ statistical test. In one case (ObsID 00031306006) the exposure was lower than expected ($1.4$~ks) and it was necessary to use the Cash statistical test (Cash 1979), which allows parameters estimation in low counts measurements through the likelihood ratio. The spectra were fitted with a power-law model with Galactic absorption ($5.22\times 10^{22}$~cm$^{-2}$, Kalberla et al. 2005) and the results are summarized in Table~\ref{tab:swift} and Fig.~\ref{fig:latxrt}.

\begin{figure}[!ht]
\centering
\includegraphics[angle=270,scale=0.7]{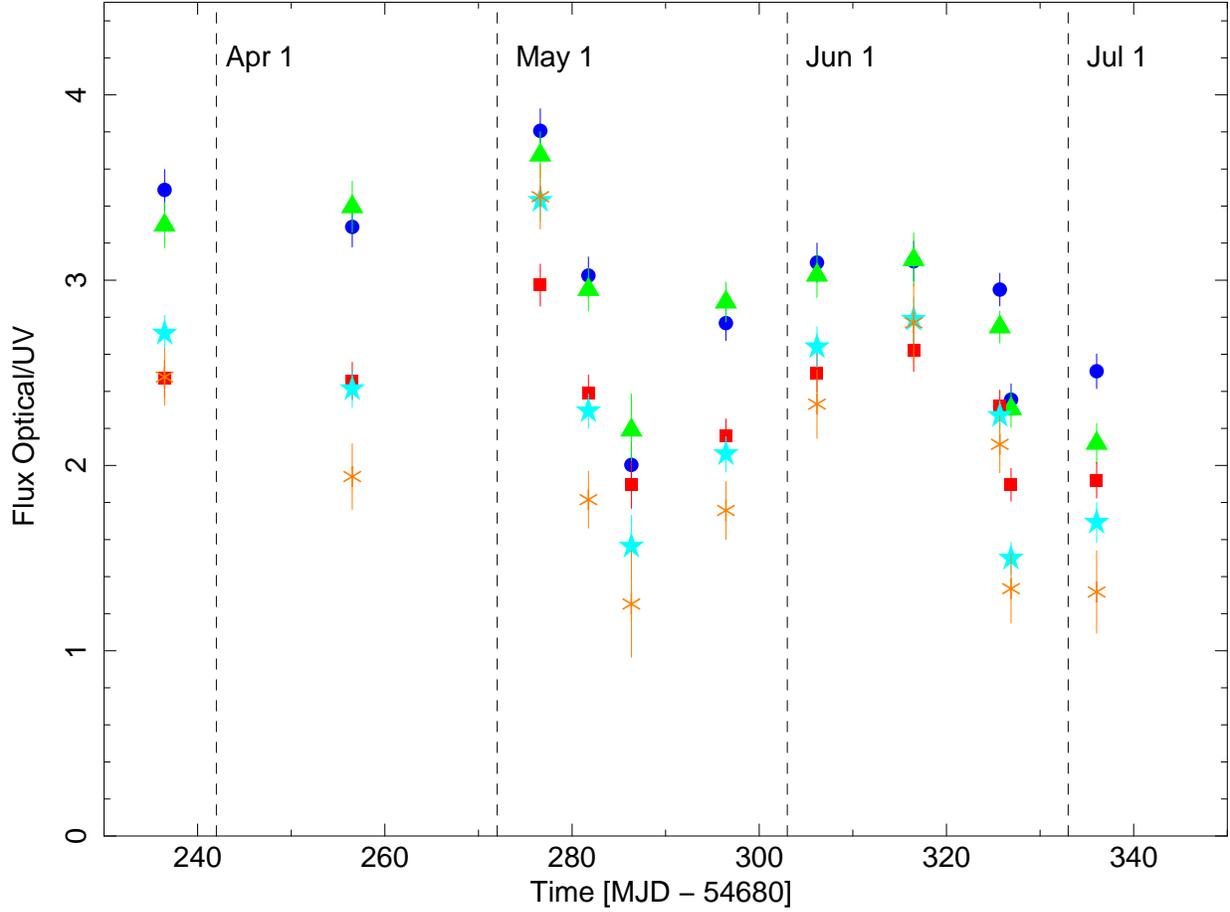}
\caption{\emph{Swift}/UVOT light curves of PMN~J0948+0022 for the three ultraviolet filters (squares: UVW1; triangles: UVM2; circles: UVW2) and two optical filters (stars: U; asterisks: V). Fluxes ($\nu F_{\nu}$) are in units of $10^{-12}$~erg~cm$^{-2}$~s$^{-1}$.}
\label{fig:uv}
\end{figure}

\begin{figure}[!t]
\centering
\includegraphics[angle=270,scale=0.7]{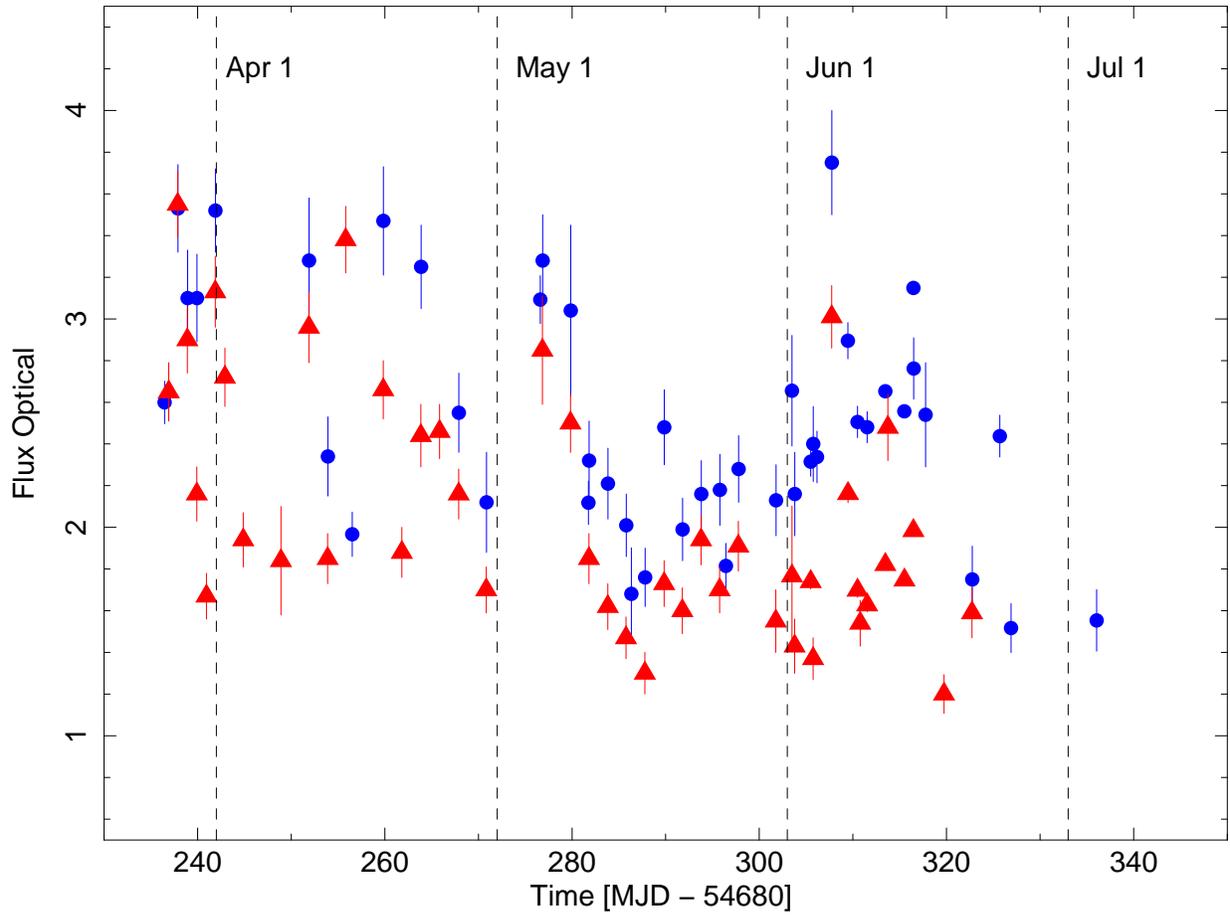}
\caption{\emph{Swift}/UVOT (B), ATOM (B, R) and SMARTS (B, R) optical light curves of PMN~J0948+0022 (circles: B; triangles: R). Fluxes ($\nu F_{\nu}$) are in units of $10^{-12}$~erg~cm$^{-2}$~s$^{-1}$.}
\label{fig:optical}
\end{figure}

UVOT counts in all the 6 available filters (V, B, U, UVW1, UVM2, UVW2) were extracted from a source region radius of $5''$ and a background region with radius $1'$, centered in a nearby source-free region and not in an annulus region around the source because of nearby contaminating sources. The observed magnitudes were corrected for the Galactic absorption $A_{V}=0.277$~mag. The absorption for the other filters was calculated according to the extinction laws of Cardelli et al. (1989). The dereddened magnitudes were converted into fluxes in physical units taking into account the zeropoints by Poole et al. (2008). Data are displayed in Fig.~\ref{fig:uv} and \ref{fig:optical}.

We note that the optical/IR filters bandpasses of the several facilites employed in this research (UVOT and the other ground-based telescopes ATOM, SMARTS, INAOE, WIRO described in the next Sections) do not match exactly. However, after a careful inspection of simultaneous or quasi-simultaneous observations, we found that these mismatches in filter bandpasses are negligible because they are smaller than the error bars.

We note also that at all the UV/optical/NIR wavelengths the quasar is unresolved with no hint of a contribution from starlight of the underlying galaxy.

\subsection{Automatic Telescope for Optical Monitoring for H.E.S.S. (ATOM)}
Optical observations in Johnson R and B filters for this campaign were obtained between March 27 and May 20 with the 0.8~m optical telescope ATOM in Namibia. ATOM is operated robotically by the H.E.S.S. collaboration and obtains automatic observations of confirmed or potential $\gamma-$bright blazars. Data analysis (debiassing, flat fielding, photometry using \texttt{SExtractor}; Bertin \& Arnouts, 1996) is conducted automatically. The magnitudes were then corrected for galactic extinction using the extinction laws of Cardelli et al. (1989), assuming $R_V=3.1$ and $A_V=0.277$~mag, which gives an absorption of $A_B=0.37$~mag and $A_R=0.23$~mag. The magnitudes were converted in fluxes using the zeropoints of Bessell (1979). Data are shown in Fig.~\ref{fig:optical}.

\subsection{Small and Moderate Aperture Research Telescope System (SMARTS)}
The source was monitored at the Cerro Tololo Inter-American Observatory (CTIO) SMARTS $1.3$~m telescope plus ANDICAM, which is a dual-channel imager with a dichroic linked to an optical CCD and an IR imager, from which it is possible to obtain simultaneous data from $0.4$ to $2.2$~$\mu$m. Optical/Near-Infrared (NIR) observations with the filters B, R and J were carried out between 2009 June 1 and 14 (MJD 54983-54996). 

\begin{figure}[!t]
\centering
\includegraphics[angle=270,scale=0.7]{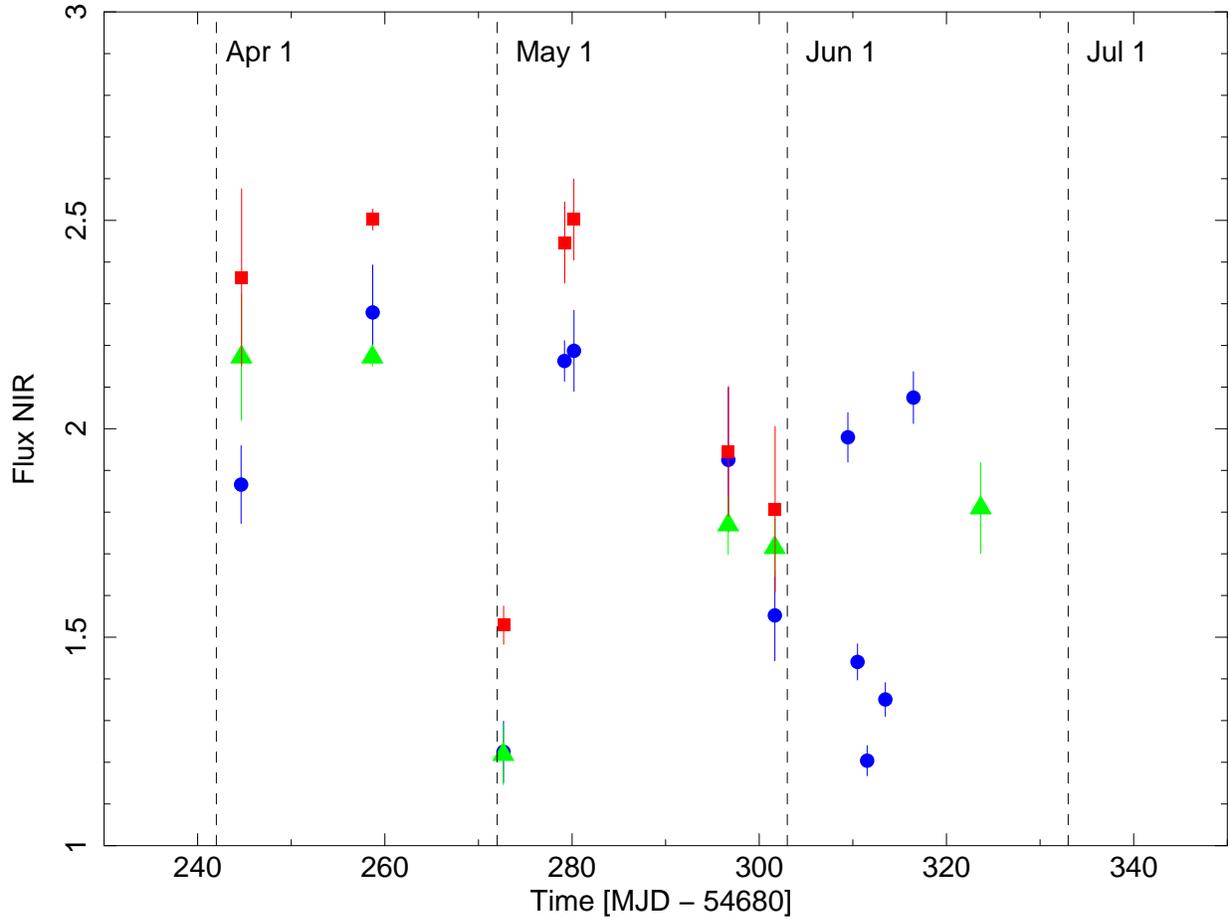}
\caption{SMARTS (J), WIRO (J, K$_{s}$) and INAOE (J, H, K$_{s}$) near-infrared light curves of PMN~J0948+0022 (circles: J; triangles: H; squares: K$_{s}$). Fluxes ($\nu F_{\nu}$) are in units of $10^{-12}$~erg~cm$^{-2}$~s$^{-1}$.}
\label{fig:nir}
\end{figure}

Optical data were bias-subtracted, overscan-subtracted, and flat-fielded using the \texttt{ccdproc} task in \texttt{IRAF}. The optical photometry was calibrated absolutely using published magnitudes (from the USNO-B1.0 catalogue) of secondary standard stars in the field of the object. IR data were sky-subtracted, flat-fielded, and dithered images combined using in-house \texttt{IRAF} scripts. The IR photometry was absolutely calibrated using 2MASS magnitudes of a secondary standard star. We estimated photometric errors by calculating the $1\sigma$ variation in magnitude of comparison stars with comparable magnitude to PMN J$0948+0022$ in the same frame. The results are summarized in Fig.~\ref{fig:optical} and \ref{fig:nir}.

\subsection{Instituto Nacional de Astrof\'isica, \'Optica y Electr\'onica (INAOE)}
NIR observations of PMN J0948+0022 were done between 2009 April 3 and June 21, at the $2.1$~m telescope ``Guillermo Haro'', with the NIR camera ``CANICA'', equipped with a Rockwell $1024 \times 1024$ pixel Hawaii infrared array, working at $75.4$~K, with standard  J(1.164 - 1.328 $\mu$m), H(1.485 - 1.781 $\mu$m) and K$_{\rm s}$ (1.944 - 2.294 $\mu$m) filters. The plate scale is $0.32$ arcsec/pix. Observations were carried out in series of $10$ dithered frames in each filter. A proper number of additional observations were adopted for the K$_s$ observations. Data sets were coadded after correcting for bias and flat-fielding. Flats were obtained from sky frames derived from the dithered ones. Data are shown in Fig.~\ref{fig:nir}.

\subsection{University of Wyoming Infrared Observatory (WIRO)}
The NIR observations at WIRO of PMN J0948+0022 were obtained on 2009 May 8-9, as part of a blazar observing campaign in which selected AGN are monitored over timescales of months, once the AGN is measured by the LAT to exceed a nominal threshold of $15 \times 10^{-8}$~ph~cm$^{-2}$~s$^{-1}$ ($E > 100$~MeV). The NIR camera is sited on the Wyoming Infrared Observatory's $2.3$~m telescope, which is optimized for IR observations, and located on Mt. Jelm at an elevation of $2943$~m. The detector is a professional grade $256^2$~InSb chip -- a spare from the \emph{Spitzer} mission -- with a square $100''$ field of view. Once accounting for atmospheric absorption, the camera's $J$ ($1.171-1.328$~$\mu$m) and $K$ ($1.987-2.292$~$\mu$m) filters have bandpasses and center wavelengths very similar to the MKO-NIR system (Tokunaga \& Vacca 2005).

The observations of PMN J0948+0022 were made on 2009 May 8-9:  sixteen $26$~s integrations each in the $J$ and $K$ filters, per night.  Each set of frames was flat-fielded and reviewed for transparency -- with maximum of three frames per set discarded -- and the remaining retained frames stacked in registration.  The source fluxes were then compared with fluxes of same or near-frame stars as well as with fluxes of NIR Arnica\footnote{see Table 2 in L. Hunt et al., Arcetri Technical Report no. 3, 1994:\\ \texttt{http://www.arcetri.astro.it/irlab/instr/arnica/arnica.html}} standards stars which were also obtained with the NIR camera, to derive $J$ and $K$ magnitudes. Data are shown in Fig.~\ref{fig:nir}.

\subsection{Owens Valley Radio Observatory (OVRO)}
PMN~J0948+0022 has been observed regularly between 2009 March 26 and July 3, at 15~GHz by the Owens Valley Radio Observatory (OVRO) 40~m telescope as part of an ongoing \emph{Fermi} blazar monitoring program.  Flux densities were measured using azimuth double switching as described in Readhead et al. (1989). The relative uncertainties in flux density result from a 5~mJy typical thermal uncertainty in quadrature with a 1.6\% non-thermal random error contribution.  The absolute flux density scale is calibrated to about 5\% using the Baars et al. model for 3C~286 (Baars et al. 1977).  This absolute uncertainty is not included in the plotted errors. The light curve is shown in Fig. \ref{fig:ovro}.

\subsection{Mets\"ahovi}
The 37~GHz observations were made between 2009 April 10 and May 30, with the 13.7~m diameter Mets\"ahovi radio telescope, which is a radome enclosed paraboloid antenna situated in Finland. A typical integration time to obtain one flux density data point is $1200-1400$~s. The detection limit of our telescope at 37~GHz is on the order of 0.2~Jy under optimal conditions. Data points with a signal-to-noise ratio $< 4$ are handled as non-detections.

The flux density scale is set by observations of DR 21. Sources 3C 84 and 3C 274 are used as secondary calibrators. A detailed description on the data reduction and analysis is given in Ter\"asranta et al. (1998). The error estimate in the flux density includes the contribution from the background and the uncertainty of the absolute calibration. The light curve is shown in Fig. \ref{fig:ovro}.

\subsection{RATAN-600}
The $2-22$~GHz instantaneous radio spectrum of PMN~J0948+0022 was observed two times, on 2009 March 24 and 25, with the 600-meter ring radio telescope RATAN-600 (Korolkov \& Parijskij 1979) of the Special Astrophysical Observatory, Russian Academy of Sciences, located in Zelenchukskaya, Russia. The broad-band radio continuum spectrum was measured quasi-simultaneously (within several minutes) in a transit mode at five different bands. Details on the method of observation, data processing, and amplitude calibration are described in Kovalev et al. (1999). The presented data were collected using the Northern ring sector of RATAN-600. Averaged flux density spectrum is presented in Fig. \ref{fig:rspec}.

\subsection{F-GAMMA (Effelsberg)}
The radio spectrum of PMN J0948+0022 at centimeter wavelength was measured with the Effelsberg 100~m telescope, within the project F-GAMMA, the monitoring program of \emph{Fermi} $\gamma$-ray blazars (F-GAMMA project, Fuhrmann et al. 2007). The observations were performed at different epochs (from 2009 April 13 to June 27), with the secondary focus heterodyne receivers between 2.64 and 42~GHz, and quasi-simultaneously with cross-scans, that is slewing over the source position, in azimuth and elevation direction, with adaptive numbers of sub-scans in order to reach the required sensitivity (for details, see Fuhrmann et al. 2008; Angelakis et al. 2008). Pointing off-set correction, gain correction, atmospheric opacity correction and sensitivity correction have been applied to the data. The results are summarized in Fig. \ref{fig:rspec}.

\subsection{Monitoring Of Jets in Active galactic nuclei with VLBA Experiments (MOJAVE)}
PMN J0948+0022 was observed on 2009 May 28 within the framework of the program MOJAVE, which is a survey with the Very Large Baseline Array (VLBA) at 15.4~GHz aiming at the study of the parsec-scale structure of relativistic jets in sources with declination $>-30^{\circ}$ (Lister et al. 2009). Total intensity and linear polarization were measured (Fig.~\ref{fig:mojave}). The total integrated flux density is $S_\mathrm{VLBA}=437$~mJy (peak value: 425 mJy/beam), while the integrated linear polarization is 3.5~mJy (peak value: 3.6~mJy/beam). The relative error in both cases is about 5\%. For details of data processing we refer to Lister et al. (2009) and Lister \& Homan (2005).

These observations were performed with a 512~Mbps recording rate and resulted in a very high dynamic range (about 8,000:1) parsec-scale total intensity image. The structure was modeled using three components with circular Gaussian intensity profiles. It was found that the VLBA core highly dominates the emission and is unresolved: the core flux density $S_\mathrm{core}=420$~mJy covers 96\% of the total parsec-scale emission. We estimated an upper limit of the core size following Kovalev et al. (2005), which turned out to be $\theta_\mathrm{core}<60$~$\mu$as (confidence level $>99$\%). We are able to get such a small upper limit, because of the high dynamic range and the simplicity of the source structure. The core brightness temperature in the source frame is estimated to be greater than $1.0\times10^{12}$~K.

The object is highly compact in comparison to the sample of radio-loud AGN reported by Lister \& Homan (2005). Its core-to-jet flux density ratio is about 25, well above the average value of 3 in the sample. However, the 0.7\,\% fractional linear polarization of the structure is in agreement with the average distribution of bright quasars (Lister \& Homan 2005).

Another MOJAVE observation was performed after the end of the campaign (2009 July 23, not shown here\footnote{See \texttt{http://www.physics.purdue.edu/astro/MOJAVE/sourcepages/0946+006.shtml}}), revealing that the flux density at 15~GHz was already decreasing ($S_\mathrm{VLBA}=340$~mJy), and the parsec-scale core appeared to be fainter than in 2009 May.

\subsection{e-VLBI}
PMN J0948+0022 was observed with the e-VLBI (electronic Very Long Baseline Interferometry) technique on April 21 at 1.6~GHz, and on May 23, Jun 10, and July 4 at 22~GHz. The epoch at 1.6 GHz was a pilot observation, lasting about 80 minutes with EVN (European VLBI Network) stations only. In the following epochs, EVN telescopes were joined by Australian and Japanese antennas, for about 9 hours at each epoch with about 1 hour of mutual visibility between Europe, Asia, and Australia (except in the last epoch).

Real time fringes were detected in all baselines between participating telescopes at all epochs. This includes Europe-Australia baselines as long as $12000$~km, which reveals that the source is highly compact and allows us to constrain its angular size. From visibility model fitting to the first $22$~GHz epoch, we determine an upper limit to the core size of 0.2~mas. This corresponds to a lower limit for the brightness temperature of $T_B > 2.9 \times 10^{10}$ K, and is consistent with the result from the second 22~GHz epoch and the 15~GHz data from MOJAVE. The 1.6~GHz observation and the final 22~GHz one lacked Europe-Australia baselines, resulting in less tight constraints. Also,
the source shows an inverted spectrum between 1.6 and 22~GHz, being only 0.17~Jy at 1.6~GHz and 0.41~Jy at 22~GHz (weighted average), with a spectral index of $-0.3$ ($S_{\nu}\propto \nu^{-\alpha}$). 

Extended emission is not revealed within our noise levels of about $1$~mJy/beam. The elongation of the fitted Gaussian is roughly along the extended emission seen at $15$~GHz, but the extended emission is resolved out in these maps. Further details on the observations and a higher level analysis will be presented in a forthcoming publication (Giroletti et al., in preparation). The results are summarized in Table~\ref{tab:eVLBI}.

\section{Spectral Energy Distributions (SEDs)}
We have built optical-to-$\gamma$ rays SED by picking time intervals so that they would be centered on the epoch of the \emph{Swift} observations (see Table~\ref{tab:swift}). We used the data from \emph{Swift} XRT and UVOT, and, when available, the optical/NIR data from ground-based facilities\footnote{Radio data were not used in the fit of the SEDs, because they are generated in regions external to that where optical-to-$\gamma$ rays are produced. More details on radio observations will be presented in Giroletti et al. (in preparation).}. In the case of $\gamma$ rays, we adopted an integration time of 5 days, centered on the day of the \emph{Swift} snapshot. The integrated LAT data were analyzed in two energy bands ($0.1-1$ and $1-10$ GeV) and we have taken as detections those with $TS \geq 9$. We have also re-built the SED corresponding to the \emph{Swift} observation performed on 2008 December 5, which was reported in Abdo et al. (2009a). However, this time, we used for LAT the data integrated over 5 days (instead of 5 months). The 12 SEDs are displayed in Fig.~\ref{fig:COMBO}. 

We have modeled these SEDs with the synchrotron and inverse-Compton (IC) model, which is described in detail in Ghisellini \& Tavecchio (2009) and was also used in the previous study (Abdo et al. 2009a). For the sake of simplicity, we just recall some basic definitions and symbols used in the present work. 

The emitting blob of plasma has spherical shape with size $r$ and is located at a distance $R_{\rm diss}$ from the central spacetime singularity with mass $M=1.5\times 10^{8}M_{\odot}$ (see Abdo et al. 2009a), moving with constant bulk Lorentz factor $\Gamma_{\rm bulk}=10$.

The energy distribution of the injected relativistic electrons has a broken power-law model, with shapes defined by $\gamma_{\rm e}^{-s_1}$ and $\gamma_{\rm e}^{-s_2}$, below and above $\gamma_{\rm e, break}$, respectively, where $\gamma_{\rm e}$ is the random Lorentz factor of electrons. This input distribution is then modified according to the radiative cooling occurring during a finite time of injection (the light crossing time of the blob) and the possibility of pair production through $\gamma\gamma \rightarrow e^{\pm}$. The distribution in output is then used to generate the observed radiation through the synchrotron, synchrotron self-Compton (SSC) and external-Compton (EC) processes. The seed photons for EC are generated directly by the accretion disk and its X-ray corona, the broad-line region (BLR), and the infrared torus. 

Obviously, in this case, the BLR emits only narrow-lines, but what is important with respect to EC is the energy density in the comoving frame. As already outlined in Abdo et al. (2009a), the differences of the BLR in NLS1s are thought to be due to (1) a disk-like shape of the BLR (Decarli et al. 2008) or (2) a shift of the BLR farther from the central supermassive singularity due to the radiation pressure of the highly accreting disk (Marconi et al. 2008). From the point of view of generating seed photons for EC, in case (1) there is no difference from a shell-like shape of the BLR, since what is important is the angle with which the blob sees the BLR (see angles $\alpha_1$ and $\alpha_2$ in Fig. 1 of Ghisellini \& Tavecchio 2009). In case (2), we performed some tests by pushing the BLR further out (up to $5\times 10^{17}$~cm), but we found minimal changes in the parameters. We note also that the size of the BLR is defined on the basis of the accretion disk luminosity, which in turn is measured from the SED, as $R_{\rm BLR}=10^{17}\sqrt{L_{\rm disk,45}}$, where $L_{\rm disk,45}$ is the luminosity of the disk in units of $10^{45}$~erg~s$^{-1}$.

The maximum electron energy is reached with $\gamma_{\rm e,max}$, and that corresponding to the IC peak is $\gamma_{\rm e,peak}$. The injected power in the form of relativistic electrons is $L_{\rm e}'$ (comoving frame), while the power carried out by the jet is composed of kinetic motion of electrons ($L_{\rm e}$) and protons ($L_{\rm p}$, one for each electron), radiation ($L_{\rm rad}$) and magnetic field ($L_{\rm B}$). 

The summary of the 12 SED fits is reported in Table~\ref{tab:ghisellini}, while Fig.~\ref{fig:model1} and \ref{fig:model2} display the evolution of some parameters on a time scale coordinated with those of the light curves of Figs.~\ref{fig:latxrt}-\ref{fig:ovro}, to allow an easy comparison with observations.

We have also built an overall SED from the averages of all the data collected in this campaign (Fig.~\ref{fig:SED}). It is not an average over the whole campaign, except for LAT data, which are collected daily. At all the other wavelengths, the result is an average of the available observations, generally limited to some periods in the campaign. The parameters obtained by the modeling of this overall SED are also reported in Table~\ref{tab:ghisellini}.

\section{Discussion}
\begin{figure}[!t]
\centering
\includegraphics[angle=270,scale=0.7]{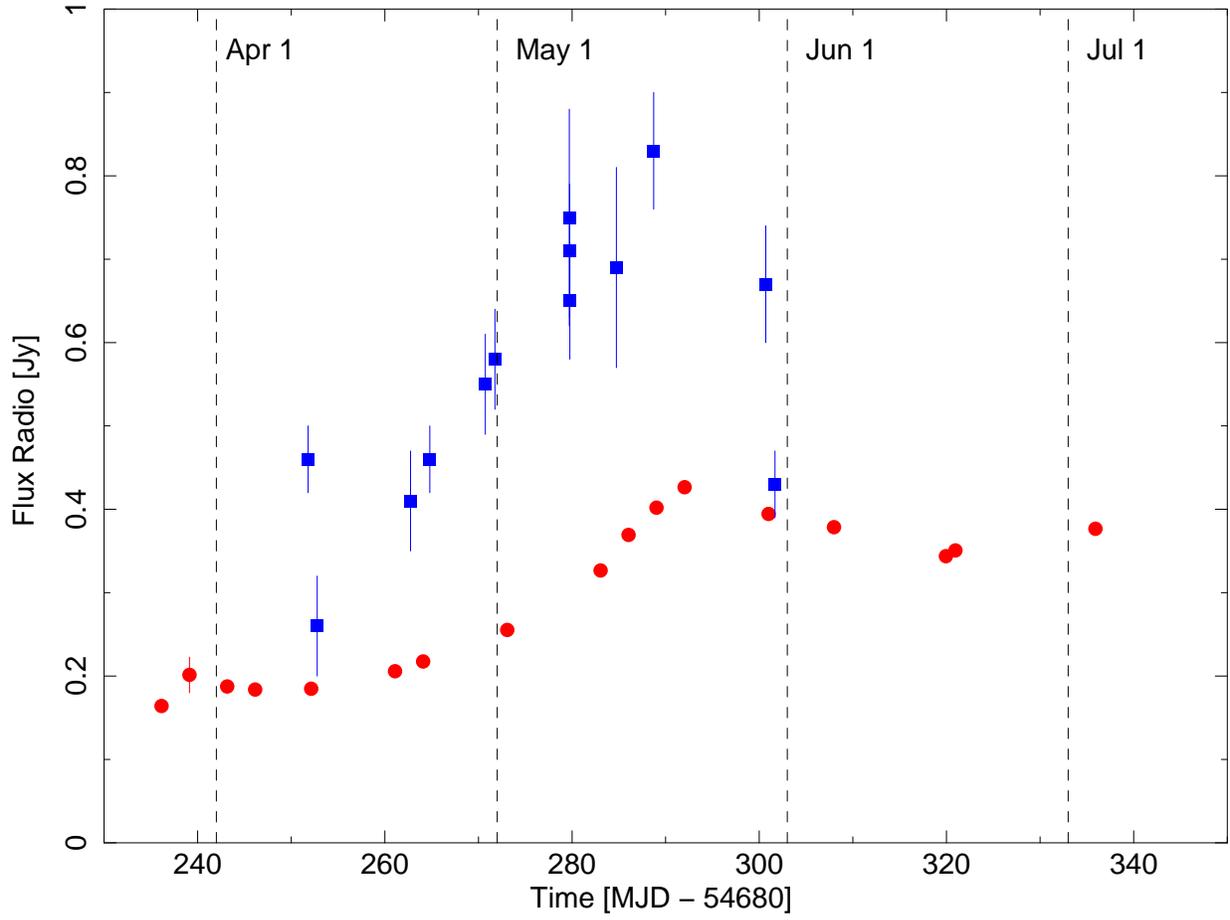}
\caption{Radio light curves of PMN~J0948+0022. Circles: 15 GHz data from OVRO; squares: 37 GHz data from Mets\"ahovi.}
\label{fig:ovro}
\end{figure}

\begin{figure}[!t]
\centering
\includegraphics[angle=270,scale=0.7]{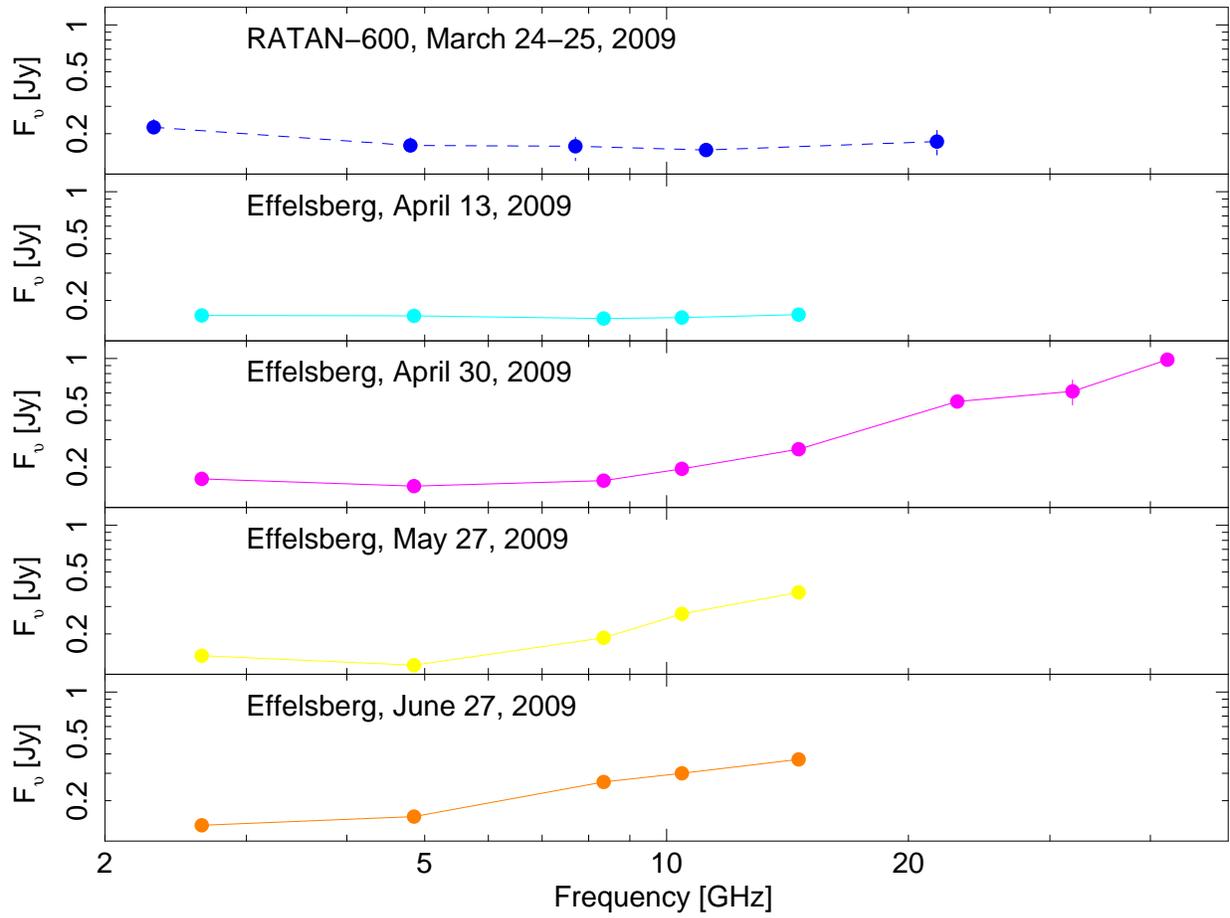}
\caption{Evolution of the radio spectrum of PMN~J0948+0022 as observed from Effelsberg and RATAN. }
\label{fig:rspec}
\end{figure}

\begin{figure}[!ht]
\centering
\includegraphics[scale=0.8]{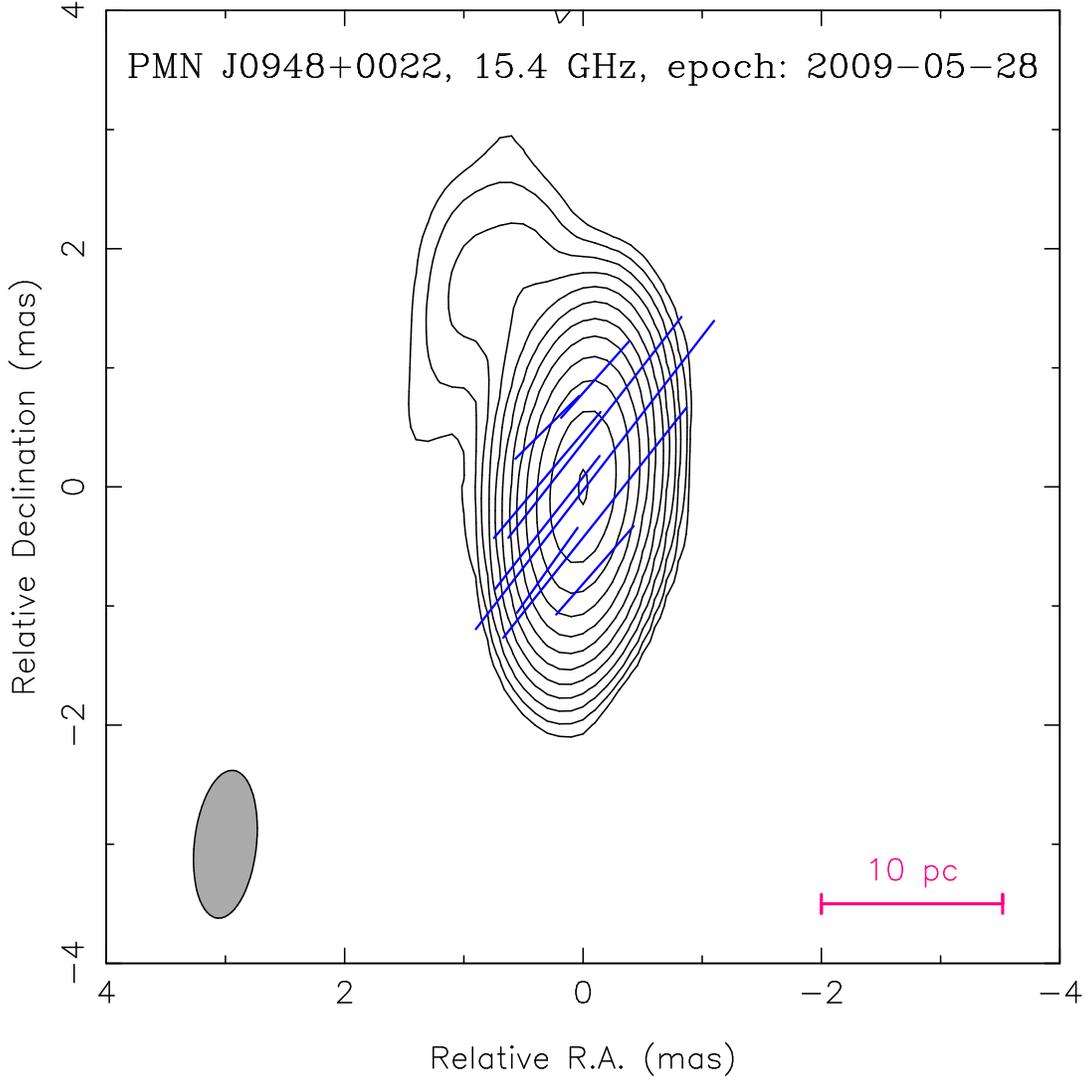}
\caption{\label{fig:mojave}
VLBA (MOJAVE program) 15~GHz combined total intensity and linear polarization image of PMN~J0948+0022 observed on 2009 May 28. The total intensity is shown by contours of equal intensity (with $\times 2$ steps). The lowest contour is $0.2$~mJy/beam and the peak intensity reaches the value of $425$~mJy/beam. The direction of the electric vectors is superimposed and represented by blue solid lines, with their length proportional to the intensity of the linear polarization, which peaks at $3.6$~mJy/beam. The FWHM size of the restoring beam is shown in the left bottom corner. The spatial scale is $6.59$~pc/mas in the adopted cosmology.}
\end{figure}

\begin{figure}[!ht]
\centering
\includegraphics[scale=0.9,clip,trim = 0 50 0 0]{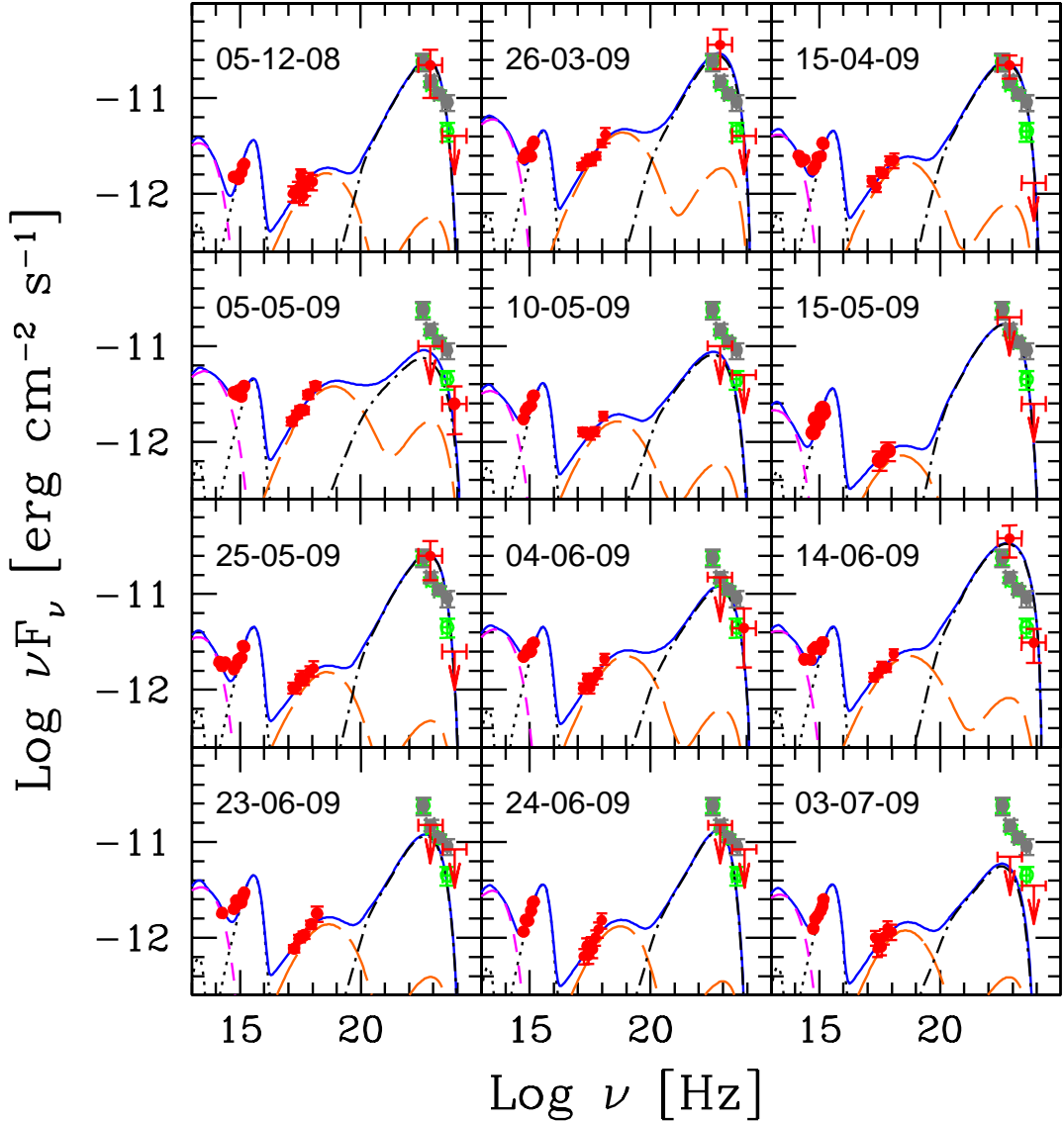}
\caption{Individual SEDs built from the measurements obtained during the observations performed in the present multiwavelength campaign and centered on the \emph{Swift} snapshots. The observation performed on December 5, 2008 is also shown (Abdo et al. 2009). Red points are the (quasi-)simultaneous data. The dotted line indicates the contribution from the accretion disk. The dashed line is the synchrotron self-Compton (SSC) and the dot-dashed line is the external Compton (EC). The blue continuous line is the sum of all the contributions. LAT spectra integrated over the three months of this campaign (grey points) and that integrated on August-December 2008 (green points) from Abdo et al. (2009) are also shown (these are almost consistent, except for the last bin at the highest energy).}
\label{fig:COMBO}
\end{figure}

\begin{figure}[!ht]
\centering
\includegraphics[angle=270,scale=0.7]{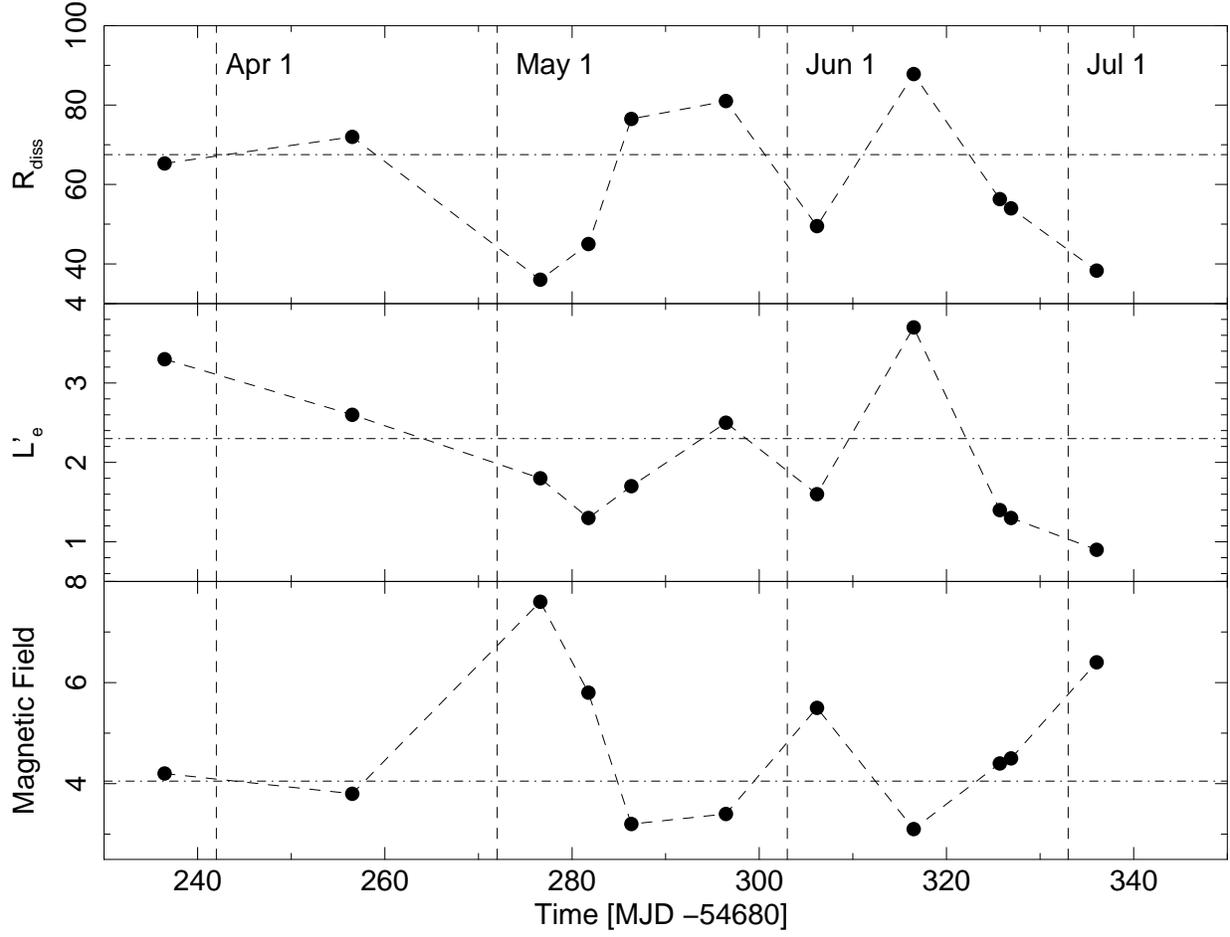}
\caption{Evolution of models parameters derived from the fits of individual SEDs. $R_{\rm diss}$ is the dissipation radius in units of [$10^{15}$~cm]; $L_{e}'$ is the injected electron power in units of [$10^{43}$~erg~s$^{-1}$]; the magnetic field $B$ is in units of [gauss]. The dot-dashed lines indicate the value obtained from the fit of the overall SED. }
\label{fig:model1}
\end{figure}

\begin{figure}[!ht]
\centering
\includegraphics[angle=270,scale=0.7]{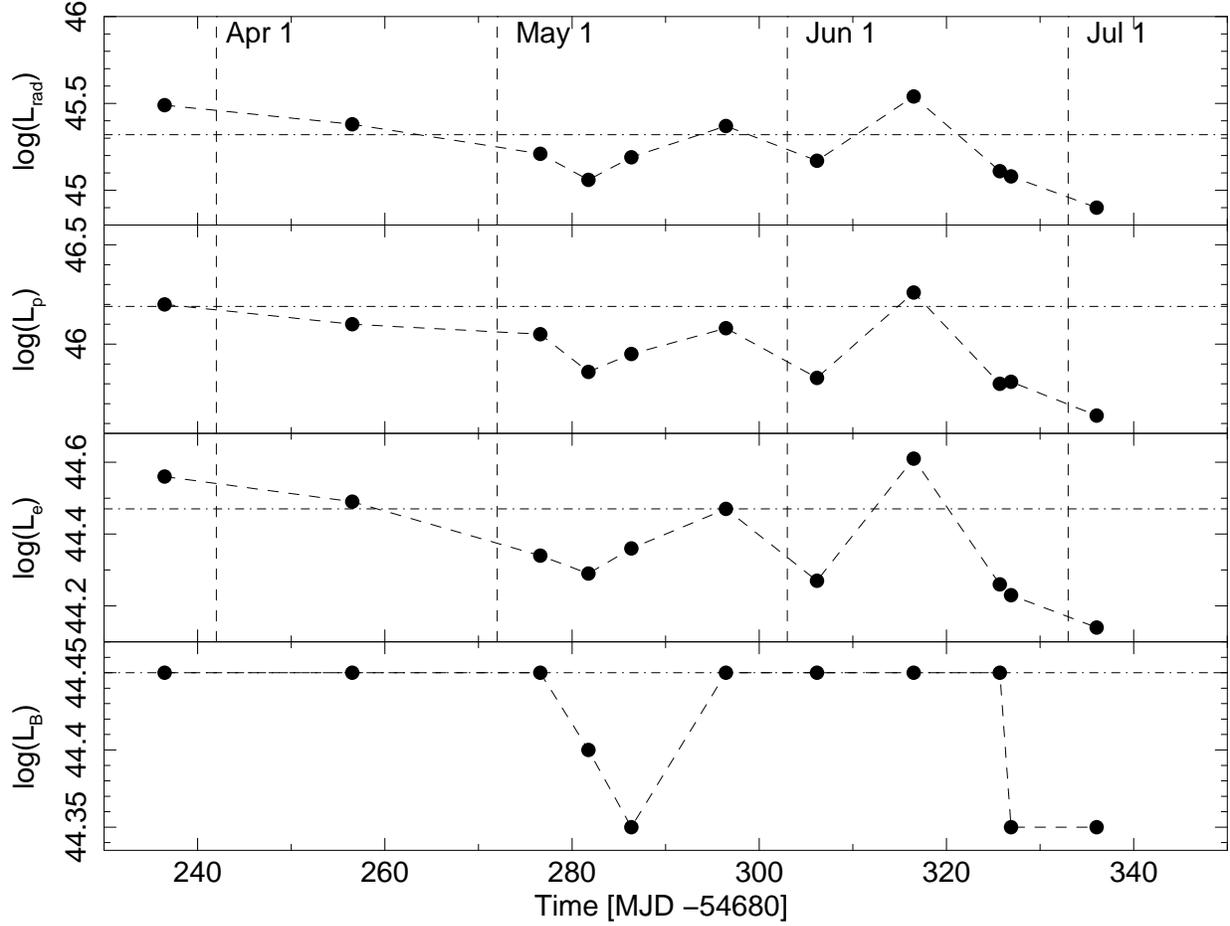}
\caption{Evolution of models parameters derived from the fits of individual SEDs. From top to bottom: radiation, proton, electron and magnetic field powers in units of [erg~s$^{-1}$]. The dot-dashed lines indicate the value obtained from the fit of the overall SED.}
\label{fig:model2}
\end{figure}

\begin{figure}[!ht]
\centering
\includegraphics[scale=0.8,clip,trim = 20 50 10 50]{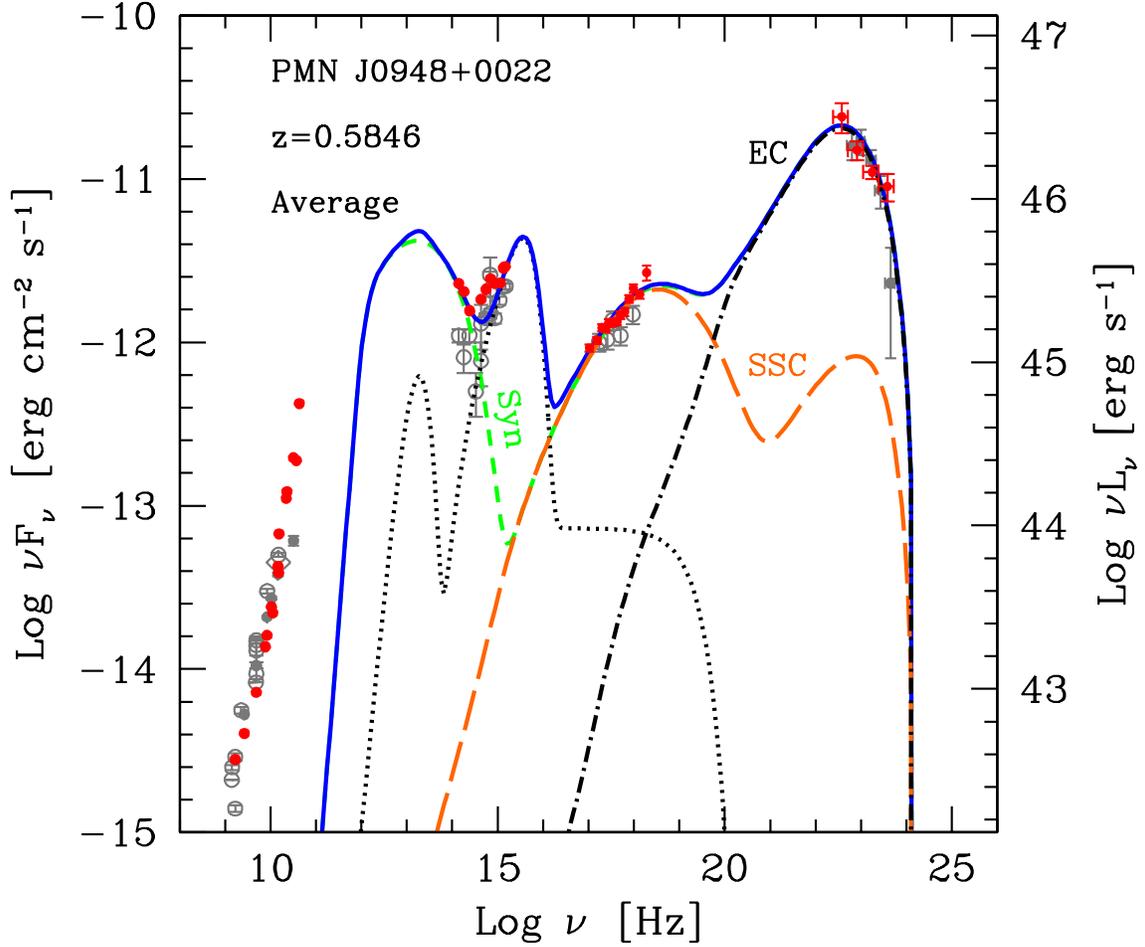}
\caption{Overall SED built from all the measurements obtained during the observations performed in the present multiwavelength campaign. Red points are the data collected during the present campaign. The dotted black line indicates the contribution from the accretion disk, X-ray corona and IR torus. The short-dashed green line is the synchrotron (Syn) and the long-dashed orange line is the synchrotron self-Compton (SSC).  The dot-dashed black line is the external Compton (EC). The blue continuous line is the sum of all the contributions. Grey symbols indicate archival data from Abdo et al. (2009). The fit does not include radio data, although they are displayed, since they are produced in regions external to that of the $\gamma$ rays. The region of synchrotron self-absorption is clearly visible around $10^{11}$~Hz.}
\label{fig:SED}
\end{figure}

\begin{figure}[!ht]
\centering
\includegraphics[angle=270,scale=0.7]{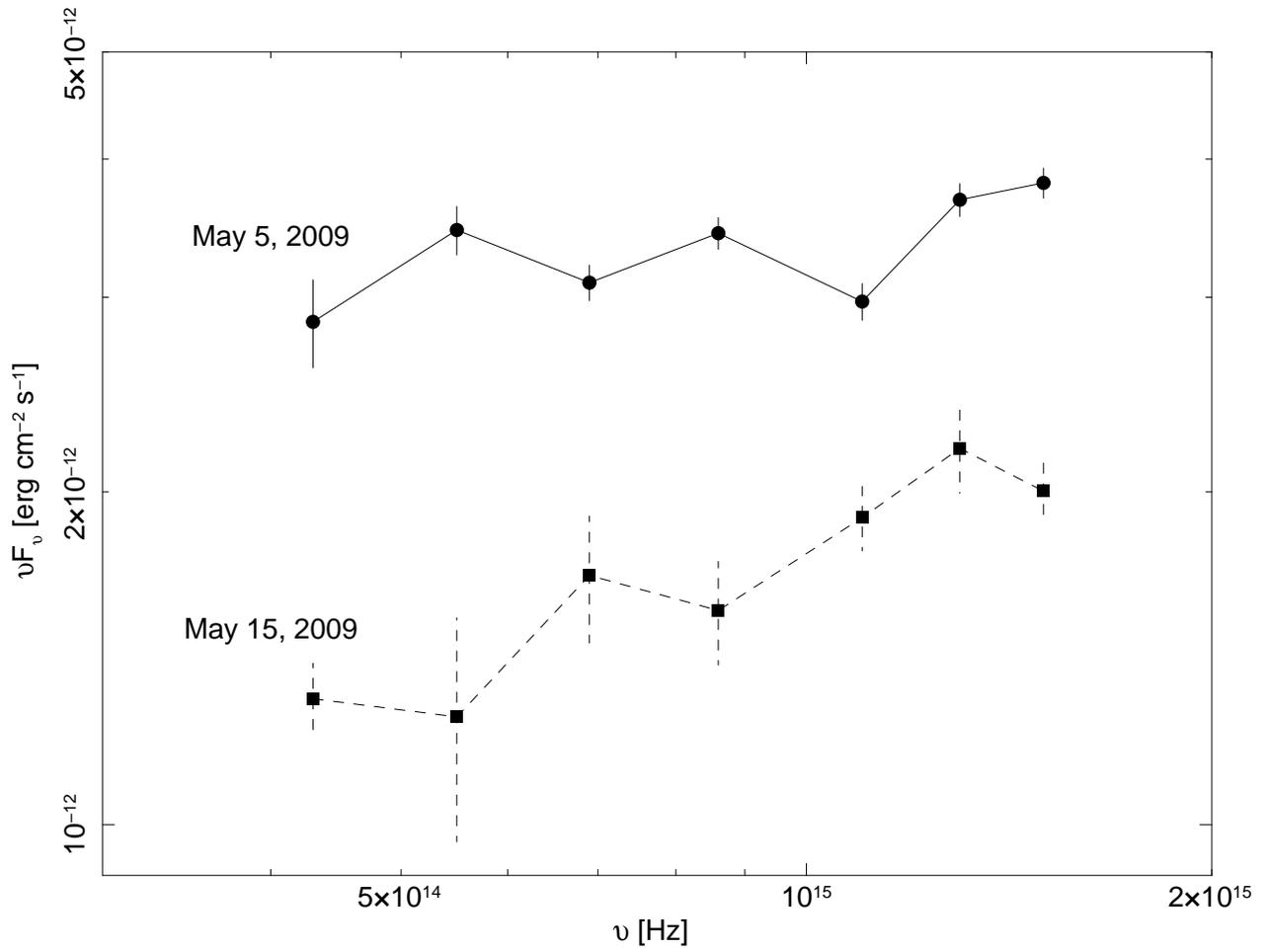}
\caption{Zoom of the SED into the optical/UV frequencies of PMN J0948+0022, as observed on May 5 and 15.}
\label{fig:uvotspec}
\end{figure}

An immediate comparison between the two average SEDs obtained from the present campaign (2009 March-July) and that of the period 2008 August-December analyzed in Abdo et al. (2009a), together with archival data (Fig.~\ref{fig:SED}), displays some changes in the emission, more pronounced at some frequencies. An inspection of the multiwavelength light curves highlights variability both in flux (at all the observed frequencies; see Fig.~\ref{fig:latxrt}-\ref{fig:ovro}) and spectral properties (Table~\ref{tab:latsummary}, Fig.~\ref{fig:rspec} and Fig.~\ref{fig:uvotspec}), except for X-rays, which show no variability in the photon index, despite showing the strongest flux variations.

To check for the presence of variability, we fitted the light curves at different frequencies with a constant flux light curve, but we got high values of reduced $\chi^2$, thus confirming that the source displayed some activity at all the wavelengths (Table~\ref{tab:variability}). The most dramatic flux changes are in X-rays (factor 3.9), radio 37 GHz (factor 3.2) and optical V and R filters (factor 2.7 and 2.9, respectively), the latter with day timescales (Table~\ref{tab:variability}). Interestingly, a clear decreasing trend from X-rays to optical wavelengths is visible at the beginning of May and corresponds to a period with decreasing $\gamma$ rays (Table~\ref{tab:latsummary}). Although only a few observations are available between May 5 and 15, the drop in flux is consistent with an exponential decay of the form $F(t)=F(t_0)\exp [-(t-t_0)/\tau]$, with a decay constant $\tau \sim 7$ days. The opposite occurs at radio and NIR frequencies, reaching a peak at 15 GHz about 20 days after the beginning of the X-ray-to-optical flux decrease. 

This coordinated trend is the typical behavior expected from the electromagnetic emission of a relativistic jet. At the radio, optical and X-ray frequencies, there is a dominance of the synchrotron and synchrotron self-Compton (SSC) processes, while $\gamma$ rays are generated by external Compton (EC). This is also clear by looking at the change in the optical/UV spectrum (Fig.~\ref{fig:uvotspec}). Indeed, it is known that these frequencies sample the rising part of the accretion disk bump, but before May 5 the optical/UV spectrum was flatter ($\alpha_{V-UVW2}=0.08\pm 0.06$) and at high fluxes(\footnote{Having defined $\alpha_{12}=-\log (F_{1}/F_{2})/\log (\nu_1/\nu_2)$, where $F_{1}$ and $F_{2}$ are the fluxes at frequencies $\nu_1$ and $\nu_2$.}). This is likely due to a higher synchrotron emission, while the accretion disk had relatively small change. In the following $\sim 10$~days, the synchrotron emission decreased to its minimum, and the shape of the optical/UV emission returned to being hard ($\alpha_{V-UVW2}=0.4\pm 0.2$) and mainly due to the rising part of the accretion disk bump. The X-ray emission followed this behavior, being due to SSC, i.e. it was high on May 5 and decreased to its minimum on May 15. 

The radio emission, coming from zones farther away from the optical-to-$\gamma$ rays dissipation region, reached its peak about 20 days after the optical-to-X-ray drop, as shown in the light curves at 15 and 37~GHz (Fig.~\ref{fig:ovro}). However, the spectral index $\alpha_{5-15\rm GHz}$, as measured between 4.85 and 14.6~GHz, changed well before, from a rather flat value ($\alpha_{5-15\rm GHz}\sim 0$) on April 13 (MJD 54934.98) and earlier, to an inverted spectrum ($\alpha_{5-15\rm GHz}=-0.40\pm 0.03$) already on April 30 (MJD 54951.75) (about two weeks, see Fig.~\ref{fig:rspec}). On May 27 (MJD 54978.79), close to the maximum flux, the spectral inversion was at its maximum too ($\alpha_{5-15\rm GHz}=-0.98\pm 0.05$) and then, on June 27 (MJD 55009.53), the spectral index was already returning to a flatter shape ($\alpha_{5-15\rm GHz}=-0.77\pm 0.04$). This is in agreement with the findings by Kovalev et al. (2009) with reference to the general radio vs $\gamma-$ray properties of the blazars detected by \emph{Fermi}/LAT during the first three months of operation (Abdo et al. 2009b). They found that the time separation between $\gamma-$ray and radio flares is typically up to a few months, in agreement with the results obtained by other authors on individual sources studies (e.g. Raiteri et al. 2008, Larionov et al. 2008, Villata et al. 2009). In the present case, if we adopt as references the peak of the $\gamma-$ray emission that occurred in the first two weeks of 2009 April and the peak of the radio flux at 15 GHz that occurred in the second half of 2009 May, we can roughly estimate a delay of 1.5-2 months.

The modeling of the SED (Fig.~\ref{fig:COMBO}, Table~\ref{tab:ghisellini}, see also the evolution of the model parameters in Fig.~\ref{fig:model1} and \ref{fig:model2}) confirmed this phenomenological view. During this campaign, the modelled values of the magnetic field, injected power, and the radius at which dissipation of energy occurs varied by factors of 2.4, 4.1 and 2.4, respectively.  At the same time, the power in radiation, electrons, protons, and the magnetic field varied by 4.4, 3, 4.2 and 1.2, respectively.  The dissipation radius was $(3.6-8.8)\times 10^{16}$~cm, roughly $0.04-0.091$~light years or $0.012-0.028$~pc from the central supermassive black hole. At the beginning of May, when the synchrotron and SSC emission dominate the optical-to-X-ray emission, the dissipation region is very compact and the magnetic field is high. The trend of the injected power (flagged by the $\gamma-$ray emission) is decreasing. Then, on May 15, the dissipation radius is larger together with a smaller value of the magnetic field. We note that the accretion remained almost constant, at about 40-50\% of the Eddington value\footnote{The Eddington value of the accretion disk luminosity corresponds to the power emitted in a condition of equilibrium between the force due to the radiation pressure and the gravity.}.

The fit from the ``overall'' SED (Table~\ref{tab:ghisellini}) had the following values: the dissipation radius is $67.5 \times 10^{15}$~cm, $L_{\rm disk} = 0.5$ times the Eddington luminosity, the injected power is $2.3 \times 10^{43}$~erg~s$^{-1}$, while the power carried out by the jet is $1.5 \times 10^{46}$~erg~s$^{-1}$ in protons, $2.9 \times 10^{44}$~erg~s$^{-1}$ in electrons, $2.1 \times 10^{45}$~erg~s$^{-1}$ in radiation, and $2.8 \times 10^{44}$~erg~s$^{-1}$ in the magnetic field. These values are well within the range of typical values for other $\gamma-$ray blazars (cf Celotti \& Ghisellini 2008, Ghisellini et al. 2009).

\section{Conclusions}
We thus confirm that PMN J0948+0022 -- despite being a radio-loud narrow-line Seyfert 1 -- hosts a relativistic $\gamma-$ray emitting jet, similar to those of FSRQs, and confirms all the hypotheses adopted to model the non-simultaneous SED in Abdo et al. (2009a). This type of source can develop a relativistic jet like blazars and radio galaxies, even though the conditions of the environment close to its central spacetime singularity are quite different. This is indeed a new class of $\gamma-$ray emitting AGN.

We have shown that the variability at multiple wavebands and the physical parameters resulting from modelling the SEDs are typical of a source midway between FSRQs and BL Lacs. The calculated powers carried by the various components of the jet are low compared to the distributions of values for FSRQ, but above those of BL Lacs (cf Celotti \& Ghisellini 2008, Ghisellini et al. 2009), and therefore within the average range of blazar powers, despite the relatively low mass of its black hole, $1.5\times 10^{8}M_{\odot}$ (Abdo et al. 2009a). The $\gamma-$ray observations performed to date have not revealed very high fluxes, i.e. above the usual threshold adopted to define an outburst in normal blazars ($F_{E>100\rm MeV}>10^{-6}$~ph~cm$^{-2}$~s$^{-1}$). However, it is not clear yet if this is due to the duty cycle of this source -- and hence if we have just observed a minor event -- or if the different environmental conditions in the core of RL-NLS1s hampers the development of a high power jet. This question will likely be answered by the continuous monitoring that \emph{Fermi}/LAT is performing on this and other sources of this type.

\acknowledgments
The \textit{Fermi} LAT Collaboration acknowledges generous ongoing support
from a number of agencies and institutes that have supported both the
development and the operation of the LAT as well as scientific data analysis.
These include the National Aeronautics and Space Administration and the
Department of Energy in the United States, the Commissariat \`a l'Energie Atomique
and the Centre National de la Recherche Scientifique / Institut National de Physique
Nucl\'eaire et de Physique des Particules in France, the Agenzia Spaziale Italiana
and the Istituto Nazionale di Fisica Nucleare in Italy, the Ministry of Education,
Culture, Sports, Science and Technology (MEXT), High Energy Accelerator Research
Organization (KEK) and Japan Aerospace Exploration Agency (JAXA) in Japan, and
the K.~A.~Wallenberg Foundation, the Swedish Research Council and the
Swedish National Space Board in Sweden. Additional support for science analysis during the operations phase is gratefully
acknowledged from the Istituto Nazionale di Astrofisica in Italy and the Centre National d'\'Etudes Spatiales in France.

This work is sponsored at PSU by NASA contract NAS5-00136.

The SMARTS observations were supported by Cycle 1 Fermi GI grant number 011283.

The Mets\"ahovi team acknowledges the support from the Academy of Finland. 

e-VLBI developments in Europe are supported by the EC DG-INFSO funded Communication Network Developments project ``EXPReS'', Contract No. 02662. The European VLBI Network is a joint facility of European, Chinese, South African and other radio astronomy institutes funded by their national research councils.

The National Radio Astronomy Observatory is a facility of the National Science Foundation operated under cooperative agreement by Associated Universities, Inc. RATAN-600 observations are supported in part by the Russian Foundation for Basic Research (projects 01-02-16812 and 08-02-00545). This research has made use of data from the MOJAVE database that is maintained by the MOJAVE team (Lister et al. 2009). The MOJAVE project is supported under National Science Foundation grant 0807860-AST and NASA-Fermi grant NNX08AV67G. 

Also based on observations with the 100-m telescope of the MPIfR (Max-Planck-Institut f\"ur Radioastronomie) at Effelsberg.

M. Hauser and S. Wagner acknowledge financial support through SFB 439 and BMBF/PT-DESY.

This research has made use of the NASA/IPAC Extragalactic Database (NED) which is operated by the Jet Propulsion Laboratory, California Institute of Technology, under contract with the National Aeronautics and Space Administration and of data obtained from the High Energy Astrophysics Science Archive Research Center (HEASARC), provided by NASA's Goddard Space Flight Center.

{\it Facilities:} \facility{ATOM (LSW)}, \facility{Effelsberg (F-GAMMA)}, \facility{e-VLBI (EVN, LBA)}, \facility{Fermi}, \facility{G. Haro (INAOE)}, \facility{Mets\"ahovi}, \facility{OVRO}, \facility{RATAN-600}, \facility{Swift},  \facility{SMARTS (Yale)}, \facility{VLBA (MOJAVE)}, \facility{WIRO}.

\clearpage

\clearpage

\clearpage

\clearpage

\clearpage

\clearpage

\clearpage

\clearpage

\clearpage

\begin{table}[!ht]
\begin{center}
\caption{Summary of the spectral fitting of the \emph{Fermi}/LAT data on a monthly timescale. \label{tab:latsummary}}
\begin{tabular}{lccc}
\tableline\tableline
Time Period & $F_{E>100 \rm MeV}$ & $\Gamma$ & TS \\
{}   &  [$10^{-7}$~ph~cm$^{-2}$~s$^{-1}$] & {} & {}\\
\tableline
April 2009 & $2.2\pm 0.4$     & $2.7\pm 0.2$ & 158\\
May 2009   & $1.2\pm 0.3$     & $2.4\pm 0.2$ & 65\\
June 2009  & $1.0\pm 0.2$     & $2.2\pm 0.2$ & 76\\
\tableline
Aug-Dec 2008 & $1.6\pm 0.1$ & $2.7\pm 0.1$ & 386\\
\tableline
\end{tabular}
\end{center}
\end{table}

\clearpage

\begin{table*}[!h]
\begin{center}
\scriptsize
\caption{Summary of results from analysis of the \emph{Swift}/XRT data. See the text for details and Fig.~\ref{fig:latxrt}.\label{tab:swift}}
\begin{tabular}{ccccccc}
\tableline
\tableline
ObsID & Time & Exposure & $\Gamma$ & Flux$_{0.2-10 \rm keV}$ & $\chi^{2}$/dof & Notes\\
{}    & [MJD]& {[ks]} & {} & [$10^{-12}$~erg~cm$^{-2}$~s$^{-1}$] & {} \\
\tableline
00031306002 & 54916.26 &  $4.8$ & $1.75\pm 0.10$ & $7.7\pm 0.3$	& $33.7/23$ & \\
00031306003 & 54936.34 & $4.4$ & $1.67\pm 0.13$ & $5.4\pm 0.3$   & $14.4/13$ & \\
00031306004 & 54956.42 & $4.8$ & $1.61\pm 0.09$ & $8.1\pm 0.3$   & $18.8/22$ & \\
00031306005 & 54961.51 & $4.9$ & $1.83\pm 0.14$ & $4.2\pm 0.4$   & $5.7/12$ & \\
00031306006 & 54966.13 & $1.4$ & $1.77\pm 0.49$ & $2.1\pm 0.7$ & $-$ & 2 PHA bins; Cash statistic (Cash 1979)\\
00031306007 & 54976.43 & $5.0$ & $1.75\pm 0.14$ & $4.3\pm 0.4$ & $7.2/11$ & \\
00031306008 & 54986.16 & $4.5$ & $1.72\pm 0.15$ & $4.3\pm 0.4$ & $9.1/10$ & \\
00031306009 & 54996.04 & $3.9$ & $1.69\pm 0.14$ & $5.4\pm 0.3$ & $14.1/11$ & \\
00031306010 & 55005.42 & $7.7$ & $1.63\pm 0.11$ & $3.6\pm 0.4$ & $7.5/15$ & \\
00031306011 & 55006.81 & $4.7$ & $1.52\pm 0.23$ & $3.3\pm 0.3$ & $3.6/6$ & \\
00038394001 & 55015.53 & $4.2$ & $1.77\pm 0.25$ & $3.1\pm 0.3$ & $3.3/5$ & \\
\tableline
\tableline
\end{tabular}
\normalsize
\end{center}
\end{table*}

\clearpage

\begin{table*}[!h]
\begin{center}
\scriptsize
\caption{Summary of the observed fluxes from e-VLBI. See the text for details.\label{tab:eVLBI}}
\begin{tabular}{ccccc}
\tableline
\tableline
Time  &  Frequency & Flux density &  $T_{B}$   &   Resolution\\
(MJD) &	  (GHz)	  & 	[Jy] &   [K] & [mas~$\times$~mas, deg]\\
\tableline
$54942$ & $1.66$  &  $0.17\pm 0.03$  &    $>1.7\times 10^{6}$ &  $35.4\times 22.9$, $12$\\
$54974$	& $22.2$	 &  $0.7\pm 0.2$    &    $>3.1\times 10^{10}$  &  $0.22\times 0.59$, $24$\\
$54992$	& $22.2$	 &  $0.3\pm 0.1$ &    $>2.3\times 10^{10}$	   &  $0.19\times 0.47$, $28$\\
$55016$	& $22.2$	 &  $0.5\pm 0.1$   &    $>1.5\times 10^{10}$	   &  $0.41\times 0.48$, $55$\\
\tableline
\tableline
\end{tabular}
\normalsize
\end{center}
\end{table*}

\clearpage

\begin{deluxetable}{ccccccccccccccc}
\tabletypesize{\scriptsize}
\rotate
\tablecaption{Summary of the fits of the SEDs. \label{tab:ghisellini}}
\tablewidth{0pt}
\tablehead{
\colhead{Time} & \colhead{$R_{\rm diss}$} & \colhead{$L_{\rm disk}$} & \colhead{$L_{\rm e}'$} & \colhead{$B$} &
\colhead{$\gamma_{\rm e,break}$} & \colhead{$\gamma_{\rm e,max}$} & \colhead{$\gamma_{\rm e,peak}$} &
\colhead{$s_1$} & \colhead{$s_2$} &
\colhead{$U'$} & \colhead{$\log L_{\rm rad}$} & \colhead{$\log L_{\rm p}$} & \colhead{$\log L_{\rm e}$} & \colhead{$\log L_{\rm B}$}\\

\colhead{(1)} & \colhead{(2)} & \colhead{(3)} & \colhead{(4)} & \colhead{(5)} &
\colhead{(6)} & \colhead{(7)} & \colhead{(8)} & \colhead{(9)} & \colhead{(10)} &
\colhead{(11)} & \colhead{(12)} & \colhead{(13)} & \colhead{(14)} & \colhead{(15)}
}
\startdata
$54916$ & 65.3 & 0.5 & 3.3 &  4.2 &  700 &  2000 & 675 &  -0.5 & 2.2 &  4.4 & 45.49 & 46.20 & 44.56 & 44.45\\
$54936$ & 72.0 & 0.5 & 2.6 &  3.8 &  600 &  1900 & 556 & -0.25 & 2.2 & 3.9 & 45.38 & 46.10 & 44.49 & 44.45\\
$54956$ & 36.0 & 0.5 & 1.8 &  7.6 &  400 &  2200 & 462 & -1.0  &  2.2 &  8.1 & 45.21 & 46.05 & 44.34 & 44.45\\
$54961$ & 45.0 & 0.45 & 1.3 &  5.8 &  500 &   1800 & 476 & 0.0 &  2.2 &  5.3 & 45.06 & 45.86 & 44.29 & 44.40\\
$54966$ & 76.5 & 0.4  & 1.7 &  3.2 &  600 &  1800 & 526 &  0.0 &  2.2 &  3.4 & 45.19 & 45.95 & 44.36 & 44.35\\ 
$54976$ & 81.0 & 0.5  & 2.5 &  3.4 & 900 &  1700 & 636 & 0.0  &  2.2 &  3.6 & 45.37 & 46.08 & 44.47 & 44.45\\
$54986$ & 49.5 & 0.5  & 1.6 &  5.5 &  700 &  2100 & 656 & -0.5 &  2.2 &  5.1 & 45.17 & 45.83 & 44.27 & 44.45\\
$54996$ & 87.7 & 0.5  & 3.7 &  3.1 &  600 &  2500 & 604 &  0.0 & 2.2 &  3.6 & 45.54 & 46.26 & 44.61 & 44.45\\
$55005$ & 56.3 & 0.5  & 1.4 &  4.4 &  600 &  1700 & 543 & -0.5 & 2.2 &  4.4 & 45.11 & 45.80 & 44.26 & 44.45\\  
$55007$ & 54.0 & 0.4  & 1.3 &  4.5 &  900 &  1400 & 640 & -1.0 &  2.2 &  4.3 & 45.08 & 45.81 & 44.23 & 44.35\\
$55015$ & 38.3 & 0.4  & 0.9 &  6.4 &  500 &  1500 & 465 & -0.5 & 2.2 &  5.8 & 44.90 & 45.64 & 44.14 & 44.35\\
\tableline
Overall & 67.5 & 0.5 & 2.3 & 4.1 & 530 & 2000 & 464 & -1.0 & 2.7 & 4.0 & 45.32 & 46.19 & 44.47 & 44.45 \\
\tableline
$54805$ & 72.0 & 0.4 & 2.3 &  3.4 &  1000 &  1500 & 623 & -0.25 & 2.2 &  3.7&  45.33 & 46.04 & 44.45 & 44.35 \\
\tableline
Abdo et al. (2009) & 67.5 & 0.4 & 3.2 & 2.4 & 800 & 1600 & 411 & 1.0 & 2.2 & 3.7 & 45.30 & 46.68 & 44.70 & 44.25\\
\enddata
\tablecomments{Columns: (1) time [MJD]; (2) radius at which most of the dissipation occurs [$10^{15}$~cm]; (3) luminosity of the accretion disk in Eddington units; (4) injected electron power in the comoving frame [$10^{43}$~erg~s$^{-1}$]; (5) magnetic field [gauss]; (6, 7, 8) random electron Lorentz factors $\gamma_{\rm e,break}$, $\gamma_{\rm e,max}$ and $\gamma_{\rm e,peak}$, respectively; (9, 10) power law indexes of the electron distribution below and above $\gamma_{\rm e,break}$, respectively; (11) radiation and magnetic energy density in the comoving frame [erg~cm$^{-3}$]; (12, 13, 14, 15) radiation, proton, electron and magnetic field power of the jet [erg~s$^{-1}$]. See the text for details and Fig.~\ref{fig:SED}.}
\end{deluxetable}

\clearpage

\begin{table}[!ht]
\begin{center}
\caption{Results of the fitting of the light curves with a constant flux line and maximum observed factor of flux change. \label{tab:variability}}
\begin{tabular}{lcc}
\tableline\tableline
Band/Filter/Frequency & $\tilde{\chi}^2$ & Factor Flux Change\\
\tableline
$\gamma$ ray & 2.0 & 2.2\\
X-ray        & 30.3 & 3.9\\
UVW2          & 22.0 & 1.9\\
UVM2          & 16.7 & 1.9\\
UVW1          & 10.2 & 1.8\\
U             & 28.3 & 2.4\\
B             & 19.6 & 2.5\\
V             & 12.5 & 2.7\\
R             & 22.4 & 2.9\\
J             & 46.8 & 1.9\\
H             & 41.6	& 1.8\\
K             & 60.9 & 1.6\\
37 GHz        & 5.9  & 3.2\\
15 GHz        & 252.3 & 2.6\\
\tableline
\end{tabular}
\end{center}
\end{table}

\end{document}